\documentclass[runningheads]{llncs}
\usepackage[T1]{fontenc}
\usepackage{graphicx,verbatim,amsfonts,amsmath}
\usepackage{booktabs}
\begin{document}
\title{GLIDE-Reg: Global-to-Local Deformable Registration Using Co-Optimized Foundation and Handcrafted Features}
\titlerunning{GLIDE-Reg}
%

\author{Yunzheng Zhu$^{1,2}$, Aichi Chien$^{2}$, Kimaya Kulkarni$^{1}$, Luoting Zhuang$^{1}$, Stephen Park$^{1}$, Ricky Savjani$^{3}$, Daniel Low$^{3}$, and William Hsu$^{1}$}  
\institute{$^{1}$ Medical \& Imaging Informatics, Department of Radiological Sciences, David Geffen School of Medicine at UCLA, USA \\
$^{2}$ Magnetic Resonance Research Labs, Department of Radiological Sciences, \\ David Geffen School of Medicine at UCLA, USA \\
$^{3}$ Department of Radiation Oncology, David Geffen School of Medicine at UCLA, Los Angeles, USA \\
    \email{yunzhengzhu19@g.ucla.edu}}
  
\maketitle              
\begin{abstract}
Deformable registration is crucial in medical imaging. Several existing applications include lesion tracking, probabilistic atlas generation, and treatment response evaluation. However, current methods often lack robustness and generalizability across two key factors: spatial resolution and anatomical coverage. To address these issues, we proposed GLIDE-Reg, a registration framework that can handle both local and large-scale deformations. We jointly optimize a registration field and a learnable dimensionality reduction module so that compressed VFM embeddings remain registration-relevant, and fuse these global semantic cues with MIND local descriptors. GLIDE-Reg achieves average dice similarity coefficients (DSC) across 6 anatomical structures of $0.859$, $0.862$, and $0.901$ in two public cohorts (Lung250M and NLST) and one institution cohort (UCLA5DCT), and outperforms the state-of-the-art DEEDS ($0.834$, $0.858$, $0.900$) with relative improvements of $3.0\%$, $0.5\%$, and $0.1\%$. For target registration errors, GLIDE-Reg achieves $1.58 mm$ on Lung250M landmarks (compared to $1.25 mm$ on corrField and $1.91 mm$ on DEEDS) and $1.11 mm$ on NLST nodule centers (compared to $1.11 mm$ on DEEDS). The substantiated performance on the nodule centers also demonstrates its robustness across challenging downstream tasks, such as nodule tracking, which is an essential prior step for early-stage lung cancer diagnosis.


\end{abstract}

\section{Introduction}
Deformable image registration (DIR) is a fundamental yet challenging task in medical imaging. The objective is to determine the optimal spatial transformation that aligns a moving image to a fixed image, typically acquired from a different viewpoint or time point. Existing clinical applications for medical image registration include longitudinal lesion tracking \cite{cai2021deep}, probabilistic atlas generation \cite{toga2001role}, radiation therapy planning \cite{lu2006deformable}, and multimodal fusion \cite{velesaca2024multimodal}. Accurate alignment of anatomical structures of interest is essential across medical imaging applications. Due to significant anatomical deformations in structures such as the lungs, heart, and vessels, DIR is often necessary. Traditional instance-optimized algorithms typically rely on image intensities or gradients, which lack semantic understanding and struggle to generalize across diverse anatomical contexts \cite{oliveira2014medical}. With the rise of deep learning, numerous DIR algorithms have achieved state-of-the-art (SOTA) performance by leveraging high-dimensional, semantically rich image representations \cite{fu2020deep}. However, deep learning-based DIR algorithms typically require extensive training and hyperparameter tuning, especially when applied to large-scale 3D medical images. Additionally, most learning-based registrations are not generalizable when adapting to a new cohort.

To address these limitations, feature-based, instance-optimized algorithms have emerged as an efficient means of leveraging semantically rich representations on a per-pair basis. Among them, ConvexAdam is a representative approach that employs the modality-independent neighborhood descriptor (MIND), a handcrafted feature capturing local voxel-level variations via 12 voxel-to-voxel distances \cite{heinrich2012mind}. Several registration algorithms have demonstrated the stability and generalizability of leveraging such feature extraction \cite{heinrich2015estimating,siebert2024convexadam,heinrich2022voxelmorph++}. Recently, with the advancement of vision foundation models (VFM), many downstream tasks have begun to leverage large pretrained models across various applications \cite{shi2023generalist,ravisam,MedSAM2}. Several existing works extracted image embeddings from the bottleneck layer of VFMs and used them as feature representations for downstream registration \cite{song2024dino,song2025dino,huang2024one,gu2025vision}, often demonstrating significant improvements in the registration of relatively large anatomical structures in the chest, such as lungs, heart, and skeleton. However, these methods do not report or demonstrate the registration of finer-structure landmarks with sufficient cases (only \cite{huang2024one} reported the performance of annotated landmarks in 20 expiration-inspiration pairs), including vessels and small pulmonary nodules, which are usually more challenging than relatively large anatomical structures and clinically critical. Thus, to ensure optimal registration of both global and local structures, we developed a global-to-local registration that aligns large-scale anatomical structures and vascular structures (Table \ref{fig:boxplot} and Fig. \ref{fig:qualres}).

One substantive issue with leveraging embeddings from VFMs is the need to compress the features to avoid the memory bottleneck and computation overhead. This is partly due to the high computational cost of processing full 3D volumes, which is already substantial even before introducing a fourth dimension (e.g, time). DINO-Reg addresses the issue using principal component analysis (PCA). However, PCA is linear and deterministic, leading to a significant loss of information. Thus, to preserve the rich semantics and enhance the transferability of VFM embeddings for registration, we proposed a dynamic dimensionality-reduction mechanism based on a variational autoencoder (VAE). The learnable parameters in the VAE can, in theory, map the entire set of VFM embeddings to a reduced space and be leveraged for registration. To avoid the VAE drifting away from registration objectives, the VAE-based dimensionality reduction is jointly updated with the registration in a dynamic fashion.

Our contributions are: (1) a co-optimized global–local registration formulation that combines semantic and spatial alignment within a single instance-specific optimization framework. GLIDE‑Reg explicitly couples foundation-model derived global semantic features with handcrafted local structural descriptors through a unified optimization pipeline; (2) a dynamic, registration-aware dimensionality reduction mechanism for VFM embeddings; (3) we demonstrate that sequentially extracted 2D VFM embeddings can be effectively repurposed for 3D deformable registration; and (4) comprehensive evaluation on heterogeneous lung CT registration tasks.


\section{Data}
\label{sec:data}
\begin{table*}[htbp]
    \caption{Specifications of Datasets used for this study.}
    \centering
    \resizebox{0.70\textwidth}{!}{
    \begin{tabular}{l|c|c|c}
    \hline
    \textbf{Dataset} & \textbf{NLST} & \textbf{Lung250M} & \textbf{UCLA5DCT} \\
    \hline
    Study Disease & 
    Indeterminate Pulmonary Nodule & COPD & $N/A$ \\
    Study Age & $55$ to $74$ & $48$ to $80$ & $N/A$ \\
    Collection Date & $1999$ to $2001$ & After $2007$ (Phase I or II) & $N/A$ \\
    
    Task (Time Diff.) & Longitudinal ($1$ to $2$ years) & Insp. Exp. ($\sim$ $30$ mins) & Free-Breathing ($<$ $5$ mins) \\
    
    Data Type & Low-Dose CT & Breath-Hold CT & Free-Breath CT \\
    
    Resolution & $512 \times 512 \times [67, 196]$ & $512 \times 512 \times [112, 135]$ & $500\times 500\times [342, 411]$ \\
    x-y px size ($mm$) & $[0.53, 0.80]$ & $[0.59, 0.74]$ & $1$ \\
    z thickness ($mm$) & $[1.8, 5.0]$ & $2.5$ & $1$ \\
    Manufacturer & GE; Philips; Siemens, etc. & GE & $N/A$ \\
    
    \# Pairs & $326$ & $10$ & $60$ \\
    \# Annotated LMs & $1$ nodule/pair & $300$/pair & $N/A$ \\
    \hline
    \end{tabular}
    }
    \label{tab0:datastats}
\end{table*}

Our study is conducted on three heterogeneous Lung CT datasets (Table \ref{tab0:datastats}). \\
\textbf{NLST} \cite{national2011reduced} consists of computed tomography (CT) scans from up to three years. 427 subjects with at least two years of CT were selected to form 1,202 pairs of scans. After data splitting, 1,018/184 pairs were used for training/validation. The test cases consist of 207 patients with indeterminate pulmonary nodules identified at least at the last timepoint. Each case was paired with one of its prior scans (as the fixed image), yielding 326 registration pairs. We also generated nodule centroids using nodule detection algorithms and confirmed with the slice number provided by NLST.
NLST were resampled to 1$mm$$\times$1$mm$$\times$1$mm$ isotropic spacing with $[252, 424] \times [252, 424] \times [250, 414]$ voxels. \\
\textbf{Lung250M-4B (Lung250M)} \cite{falta2023lung250m} is a dataset that reuses the preprocessed 10 pairs from DIR-LAB COPDgene. 
In each 3D volume pair, 300 manually annotated landmarks within the lungs are provided. Each 3D volume is already preprocessed in 1$mm$$\times$1$mm$$\times$1$mm$ isotropic spacing and $[241, 302] \times [179, 232] \times [253, 325]$ voxels. \\
\textbf{UCLA5DCT} \cite{lauria2025retrospective} is a subset of the complete UCLA5DCT dataset collected in our institution. The dataset consists of pairs of 3D Lung CT scans with breathing traces. 
The complete test dataset consists of 146 patients. Each patient has 25 fast helical free-breathing 3D scans. We randomly selected four patients, then selected six scans from each patient to form 60 pairs. 

\section{Method}
\label{sec:methods}
\begin{figure}
\includegraphics[width=\textwidth]{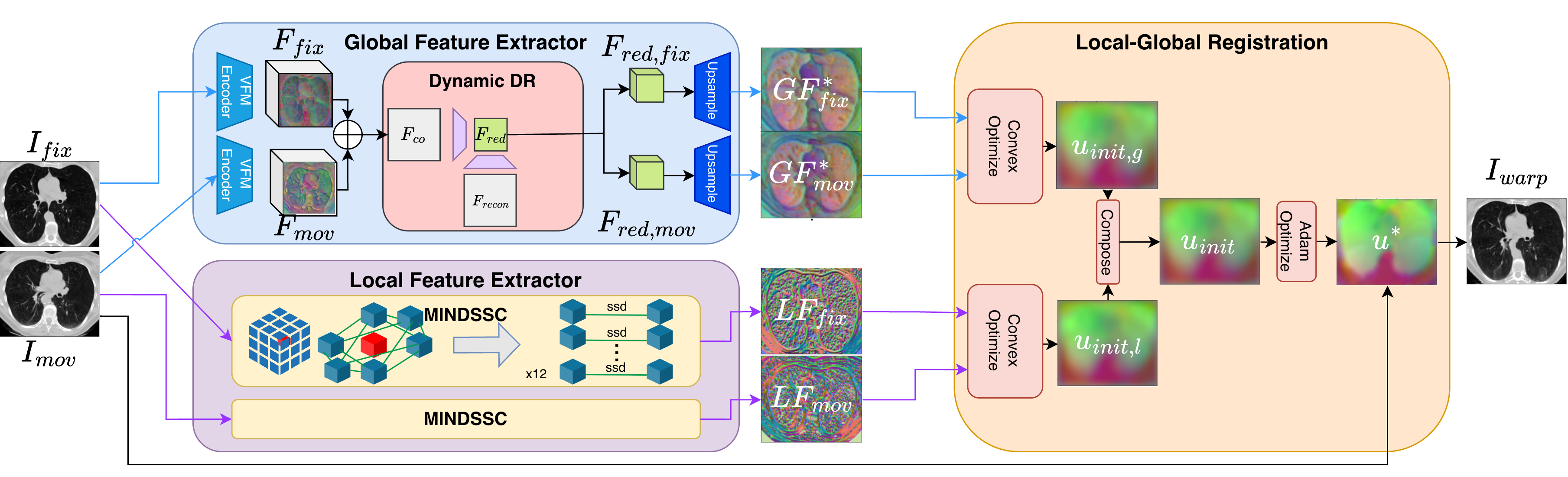}
\caption{GLIDE-Reg Pipeline. A pair of 3D images $I_{fix}$ and $I_{mov}$ are passed to a global feature extractor and a local feature extractor (MIND) for extracting the optimal global feature representations $GF_{fix}^*$, $GF_{mov}^*$ and local feature representations $LF_{fix}$, $LF_{mov}$, respectively. Then the features are registered with 1) feature-independent coupled convex optimization and 2) fused Adam optimization in Local-Global Registration and output the $I_{warp}$. DR stands for dimensionality reduction.} \label{fig:pipeline}
\end{figure}
\subsection{Problem Formulation}
Deformable image registration is commonly designed as a variational optimization problem \cite{haskins2020deep}. Assume we have a fixed image $I_{fix}: \Omega_{fix} \to \mathbb{R}^{H \times W \times D}$ and a moving image $I_{mov}: \Omega_{mov} \to \mathbb{R}^{H \times W \times D}$ within the spatial domain $\Omega \subset \mathbb{R}^{n}$. The goal is to find the spatial transformation $\varphi^{-1}: \Omega_{mov} \to \Omega_{fix}$ such that $\varphi^{-1} = I_{d}+u$, where $I_{d}$ is the identity transformation grid, and $u$ is the displacement field. We minimize the following energy:
\begin{equation}
    \hat{u} = \arg\min_{u} D(I_{fix}, I_{mov} \circ (I_{d}+u)) + \lambda r(u).
\label{eq:main}
\end{equation}
Here, $D(\cdot, \cdot)$ and $r(\cdot)$ represent the similarity and the regularization terms.
$\lambda$ is the weight for balancing the $I_{fix}$, $I_{mov} \circ \varphi^{-1}$ image pair matching and the regularization on the displacement $u$.

\subsection{Feature Extraction} 
Since we try to maintain the finer spatial details of the deep feature representation as well as the coarse structures (which are usually critical for nodule-level semantics), different from \cite{song2024dino,song2025dino,gu2025vision}, we extract 2D features maps from Segment Anything Model 2 (SAM2) \cite{ravisam} encoder ($E$), denoted as $F_{fix}^z$, $F_{mov}^z \in \mathbb{R}^{h_e \times w_e \times d}$, for all available $Z$ foreground 2D axial slices, denoted as $\{I_{fix}\}^{Z-1}_{z=0}$ and $\{I_{mov}\}^{Z-1}_{z=0}$, where $h_e$ and $w_e$ are the height and width of the VFM image embedding ($h_e = w_e = 64$), d is the embedding dimension of the feature maps ($d = 256$), and z is the slice number. Benefiting from the memory attention design in SAM2, previous frames are stored and reused in subsequent frames, greatly reducing computational complexity when extracting features from a relatively long sequence of 2D images within 30 seconds.
The 2D feature maps $F^{z} \in \mathbb{R}^{h_e\times w_e\times d}$ are concatenated into a complete 3D feature map $F \in \mathbb{R}^{h_e\times w_e\times Z\times d}$ along the axial axis.

The modality-independent neighborhood descriptor (MIND) is also extracted from each 3D image. The 12-channel feature representations are extracted, denoted as $LF_{fix}$ and $ LF _ {mov} $, with distinctive structures in the local neighborhoods of the corresponding images, $ LF_{fix}, LF_{mov} \in \mathbb{R}^{H \times W \times D \times 12}$.

\subsection{Dynamic Dimensionality Reduction}
To reduce computational cost at registration, the extremely large embedding dimension of the feature maps (256 for SAM2 and 1024 for DINO-v2-large) often necessitates dimensionality reduction. Linearly mapping from a high-dimensional embedding (e.g., 256 for SAM2) to a much lower-dimensional space (e.g., 12) can result in significant loss of semantic information, especially if the relationships are non-linear. PCA is a deterministic matrix decomposition method that computes the eigenvectors of the input covariance matrix, providing better interpretability, but inherently loses the information from the unselected features. Thus, we developed a neural network (NN)-based dimensionality reduction method to enhance semantic preservation in full-rank embeddings. 


The architecture of the dynamic dimensionality reduction mechanism is a variational autoencoder (VAE). The $(F_{fix}, F_{mov}$) feature pair is flattened and concatenated in the channel dimension to form the combined feature $F_{co}^{n\times d}$, where $n = H \times W \times D$. This is then passed to the VAE and output as $F_{recon}$, which is forced to maintain fidelity with $F_{co}$ by a mean squared error term. A Kullback-Leibler (KL) divergence is also added to regularize the latent space of VAE under a Gaussian distribution. The weights for balancing the reconstruction and KL terms are $\delta_1$ and $\delta_2$. The mean vector, which is inherent with less noise variations, in the latent features will be extracted as the dimension-reduced feature $F_{red}^{n\times 12}$, and then processed (split and upsample) into global feature pairs $(GF_{fix}, GF_{mov})$. Two optimizers are designed for joint learning of the dimensionality reduction and registration. We aim to obtain the optimal global feature pairs $(GF_{fix}^{*}, GF_{mov}^{*}) \in \mathbb{R}^{H \times W \times D \times 12}$ with an optimally refined displacement fields $u^{*}$.

\subsection{Global-to-Local Registration}
We first perform the coupled convex discrete optimization, following \cite{heinrich2014non}, on global feature pairs ($GF_{fix}^{*}$, $GF_{mov}^{*}$) and local feature pairs ($LF_{fix}$, $LF_{mov}$) independently. The resulting displacements of the convex optimization on both the global feature pairs ($\hat{u}_{g}$) and the local feature pairs ($\hat{u}_{l}$) are composed as the initial displacement, $u_{init}(X) = \hat{u}_{g}(X) + \hat{u}_{l}(X + \hat{u}_{g}(X))$ where $X$ represents a 3D coordinate, for the Adam instance-optimization.
At Adam instance-optimization, $u_{init}$ is refined over an arbitrary number of iterations to satisfy Eq. \eqref{eq:main}. Our similarity term $D(\cdot, \cdot)$ is developed for both local and global feature pairs as:
\begin{equation}
    D = \alpha L_{global}(GF_{fix}, GF_{mov} \circ (I_{d}+u) + \beta L_{local}(LF_{fix}, LF_{mov} \circ (I_{d}+u))
\end{equation}
Both $L_{global}$ and $L_{local}$ adopt the sum squared distance as the metric, and penalized with a bending energy loss $r(u) = \lambda||\nabla^2 u||^2$. $\alpha$, $\beta$, and $\lambda$ are the hyperparameters to balance the weights. 



\section{Experiments}
\subsection{Implementation Details and Evaluation Metrics}
All experiments were performed on a single NVIDIA RTX 8000 GPU.
For all learning-based methods, we trained for 300 epochs on NLST and validated every 5 epochs. The model with the best validation loss is selected and directly evaluated on all datasets to assess cross-cohort generalization. All instance-optimization methods optimized for 800 iterations, with a learning rate of 1 and a downsampling factor of 2, following \cite{siebert2024convexadam}. For dynamic dimensionality reduction, the learning rate is set to 0.001. We empirically choose loss weights as $(\lambda$, $\alpha$, $\beta$, $\delta_{1}$, $\delta_{2}) = (1.25, 0.5, 3.5, 7.5 \times 10^4, 20)$. 

Registration performance is evaluated using the Dice Score Coefficient (DSC) on the six generated segmentation masks from TotalSegmentator \cite{wasserthal2023totalsegmentator} (lung, heart, skeleton, airway, liver, and lung vessels), and the target registration error (TRE) on the landmarks (Lung250M) and nodules (NLST). A nonpositive Jacobian determinant ($\mathbf{\%\lvert J \rvert_{<0}}$) is also evaluated on the topology preservation of the displacement fields. In addition, the average running time is reported on the same hardware.

\begin{table*}[t!]
 \caption{Comparison of registration performance on three datasets evaluated by DSC($\uparrow$) and TRE($\downarrow$) (in $mm$), and nonpositive Jacobian determinant $\%\lvert J \rvert_{<0}$ ($\downarrow$). No Reg. represents the baseline performance for the image pairs. Methods with top two performance for TRE and DSC in each column are \textbf{bolded} and \underline{underlined}, respectively. Statistical testings are computed per pair on the TRE and average DSC using Wilcoxon signed-rank test and adjusted by Benjamini-Hochberg correction with false discovery rate of $0.05$. Methods that are outperformed by or outperform GLIDE-Reg with statistical significance ($p < 0.05$) are marked with $*$ and \textdagger, respectively. }
   
    \resizebox{\textwidth}{!}{
         \begin{tabular}{lcccccccccc}
             \toprule
             & 
             \multicolumn{4}{c}{\bfseries Lung250M} & 
             \multicolumn{4}{c}{\bfseries NLST} & 
             \multicolumn{2}{c}{\bfseries [blinded]5DCT} \\
             \cmidrule(lr){2-5}
             \cmidrule(lr){6-9}
             \cmidrule(lr){10-11}
             \bfseries Method & 
              
              \bfseries TRE $\downarrow$ & 
              \bfseries DSC $\uparrow$ &
              $\mathbf{\%\lvert J \rvert_{<0}}$ $\downarrow$ &
              \textbf{Runtime} &
              
              \bfseries TRE $\downarrow$ & 
              \bfseries DSC $\uparrow$ & 
              $\mathbf{\%\lvert J \rvert_{<0}}$ $\downarrow$ &
              \textbf{Runtime} &
              
              \bfseries DSC $\uparrow$ & 
              $\mathbf{\%\lvert J \rvert_{<0}}$ $\downarrow$
              \\
              & Landmarks & Mean6 & & & Nodule & Mean6 & & & Mean6 & \\
              \midrule
              No Reg. & 
              $16.25 \pm 3.84$\textsuperscript{*} & 
              $0.517 \pm 0.080$\textsuperscript{*} & 
              $-$ &
              $-$ &

              $18.93 \pm 10.27$\textsuperscript{*} &
              $0.472 \pm 0.100$\textsuperscript{*} &
              $-$ &
              $-$ &

              $0.767 \pm 0.049$\textsuperscript{*} &
              $-$ \\
              
              \midrule
              Affine &
              $30.36 \pm 12.12$\textsuperscript{*} &
              $0.488 \pm 0.106$\textsuperscript{*} &
              $-$ &
              $<5$ sec& 
              
              $30.36 \pm 23.88$\textsuperscript{*} &
              $0.726 \pm 0.045$\textsuperscript{*} &
              $-$ &
              $<10$ sec &
              
              $0.751 \pm 0.050$\textsuperscript{*} &
              $-$ \\
              
              SyN \cite{avants2008symmetric} &
              $13.95 \pm 3.63$\textsuperscript{*} &
              $0.634 \pm 0.085$\textsuperscript{*} &
              $0.0$ &
              $< 1$ min&
              
              $17.65 \pm 10.24$\textsuperscript{*} &
              $0.534 \pm 0.125$\textsuperscript{*} &
              $< 0.001$ &
              $<2$ min &
              
              $0.838 \pm 0.036$\textsuperscript{*} &
              $< 0.001$ \\

              NiftyReg \cite{modat2010fast} &
              $14.15 \pm 3.93$\textsuperscript{*} &
              $0.563 \pm 0.086$\textsuperscript{*} &
              $0.565 \pm 0.006$ &
              $<2.5$ min&
              
              $26.93 \pm 14.25$\textsuperscript{*} &
              $0.396 \pm 0.088$\textsuperscript{*} &
              $1.846 \pm 1.992$ & 
              $<4$ min &
              
              $0.807 \pm 0.032$\textsuperscript{*} &
              $0.010 \pm 0.013$ \\

              \midrule
              VoxelMorph \cite{balakrishnan2019voxelmorph} &
              $18.09 \pm 4.64$\textsuperscript{*} & 
              $0.648 \pm 0.083$\textsuperscript{*} & 
              $1.215 \pm 0.978$ &
              $<5$ sec &

              $3.87 \pm 4.66$\textsuperscript{*} & 
              $0.807 \pm 0.042$\textsuperscript{*} & 
              $2.275 \pm 1.848$ & 
              $<5$ sec &
              
              $0.818 \pm 0.027$\textsuperscript{*} & 
              $0.284 \pm 0.293$ \\

              TransMorph \cite{chen2022transmorph} &
              $26.63 \pm 4.25$\textsuperscript{*} & 
              $0.599 \pm 0.068$\textsuperscript{*} & 
              $8.942 \pm 1.535$ &
              $<5$ sec &
              
              ${5.59 \pm 4.83}$\textsuperscript{*} & 
              ${0.772 \pm 0.041}$\textsuperscript{*} &
              ${5.024 \pm 2.680}$ & 
              $<5$ sec &
              
              ${0.814 \pm 0.027}$\textsuperscript{*} & 
              ${0.708 \pm 1.085}$ \\

              ULAE-Net \cite{shu2021medical} &
              ${16.00 \pm 3.37}$\textsuperscript{*} & 
              ${0.636 \pm 0.042}$\textsuperscript{*} & 
              ${4.433 \pm 0.987}$ &
              $<5$ sec &
              
              ${7.27 \pm 6.65}$\textsuperscript{*} &
              ${0.751 \pm 0.056}$\textsuperscript{*} &
              ${4.167 \pm 2.127}$ & 
              $<5$ sec &
              
              ${0.848 \pm 0.018}$\textsuperscript{*} &
              ${0.296 \pm 0.236}$ \\
        
              NICE-NET \cite{meng2022non} &
              ${14.04 \pm 2.93}$\textsuperscript{*} & 
              ${0.701 \pm 0.048}$\textsuperscript{*} & 
              ${0.373 \pm 0.265}$ &             
              $<5$ sec &
              
              ${3.12 \pm 3.93}$\textsuperscript{*} & 
              ${0.821 \pm 0.038}$\textsuperscript{*} &
              ${1.376 \pm 1.099}$ &
              $<5$ sec &
              
              ${0.834 \pm 0.023}$\textsuperscript{*} &
              ${0.072 \pm 0.060}$ \\

              NICE-Trans \cite{meng2023non} &
              ${18.57 \pm 5.59}$\textsuperscript{*} & 
              ${0.661 \pm 0.065}$\textsuperscript{*} & 
              ${4.029 \pm 1.108}$ &
              $<5$ sec &
              
              ${2.93 \pm 3.82}$\textsuperscript{*} & 
              ${0.818 \pm 0.043}$\textsuperscript{*} &
              ${2.100 \pm 1.932}$ & 
              $<5$ sec &

              ${0.855 \pm 0.015}$\textsuperscript{*} &
              ${0.037 \pm 0.032}$ \\

              RP-Net \cite{ma2024iirp} &
              ${16.82 \pm 4.36}$\textsuperscript{*} & 
              ${0.689 \pm 0.054}$\textsuperscript{*} & 
              ${0.272 \pm 0.261}$ &
              $<5$ sec &

              ${2.82 \pm 4.15}$\textsuperscript{*} &
              ${0.829 \pm 0.039}$\textsuperscript{*} &
              ${1.071 \pm 0.986}$ &
              $<5$ sec &

              ${0.844 \pm 0.022}$\textsuperscript{*} &
              ${0.007 \pm 0.007}$ \\
                
              \midrule
              corrField \cite{heinrich2015estimating} & 
              $\mathbf{{1.25\pm 0.27}}$\textsuperscript{\textdagger} & 
              ${0.740 \pm 0.054}$\textsuperscript{*} &
              ${0.089 \pm 0.130}$ &
              $<15$ sec&
              
              $\underline{{1.57 \pm 3.92}}$\textsuperscript{*} &
              ${0.801 \pm 0.055}$\textsuperscript{*} &
              ${0.054 \pm 0.243}$ &
              $<20$ sec &
              
              ${0.829 \pm 0.031}$\textsuperscript{*} &
              ${< 0.001}$ \\
              
              DEEDS \cite{6471238} & 
              ${1.91 \pm 0.62}$\textsuperscript{*} & 
              $\underline{{0.834 \pm 0.022}}$\textsuperscript{*} & 
              ${0.001 \pm 0.003}$ &
              $<3$ min&
              
              $\mathbf{{1.11 \pm 2.10}}$ &
              $\underline{{0.858 \pm 0.036}}$\textsuperscript{*} &
              ${0.076 \pm 0.243}$ &
              $<8$ min &

              $\underline{{0.900 \pm 0.011}}$\textsuperscript{*} &
              ${< 0.001}$ \\
              
              ConvexAdam \cite{siebert2024convexadam}& 
              ${3.25 \pm 1.30}$\textsuperscript{*} & 
              ${0.828 \pm 0.028}$\textsuperscript{*} &
              ${0.189 \pm 0.109}$ &
              $<10$ sec&
              
              ${3.06 \pm 7.50}$\textsuperscript{*} &
              ${0.827 \pm 0.089}$\textsuperscript{*} &
              ${0.489 \pm 0.506}$ &
              $<30$ sec &

              $\underline{{0.900 \pm 0.012}}$\textsuperscript{*} &
              ${0.001 \pm 0.002}$ \\
              
              DINO-Reg \cite{song2024dino,song2025dino}& 
              ${11.01 \pm 2.38}$\textsuperscript{*} & 
              ${0.774 \pm 0.018}$\textsuperscript{*} & 
              ${1.401 \pm 0.503}$ &
              $<10$ min&

              ${8.47 \pm 8.36}$\textsuperscript{*} &
              ${0.786 \pm 0.033}$\textsuperscript{*} &
              ${2.528 \pm 2.418}$ &
              $<8$ min &

              ${0.846 \pm 0.020}$\textsuperscript{*} &
              ${0.045 \pm 0.036}$
              \\
              \midrule
              GLIDE-Reg (Ours) & 
              $\underline{{1.58 \pm 0.50}}$ & 
              $\mathbf{{0.859 \pm 0.013}}$ & 
              ${0.832 \pm 0.448}$ &
              $<1.5$ min&

              $\mathbf{{1.11 \pm 2.48}}$ &
              $\mathbf{{0.862 \pm 0.033}}$ &
              ${1.797 \pm 1.655}$ &
              $<3.5$ min &

              $\mathbf{{0.901 \pm 0.010}}$ &
              ${0.009 \pm 0.020}$ \\
              \bottomrule
         \end{tabular}
    }
\label{tab:results}
\end{table*}

\begin{figure}[htbp]
\centering
\includegraphics[width=0.95\textwidth]{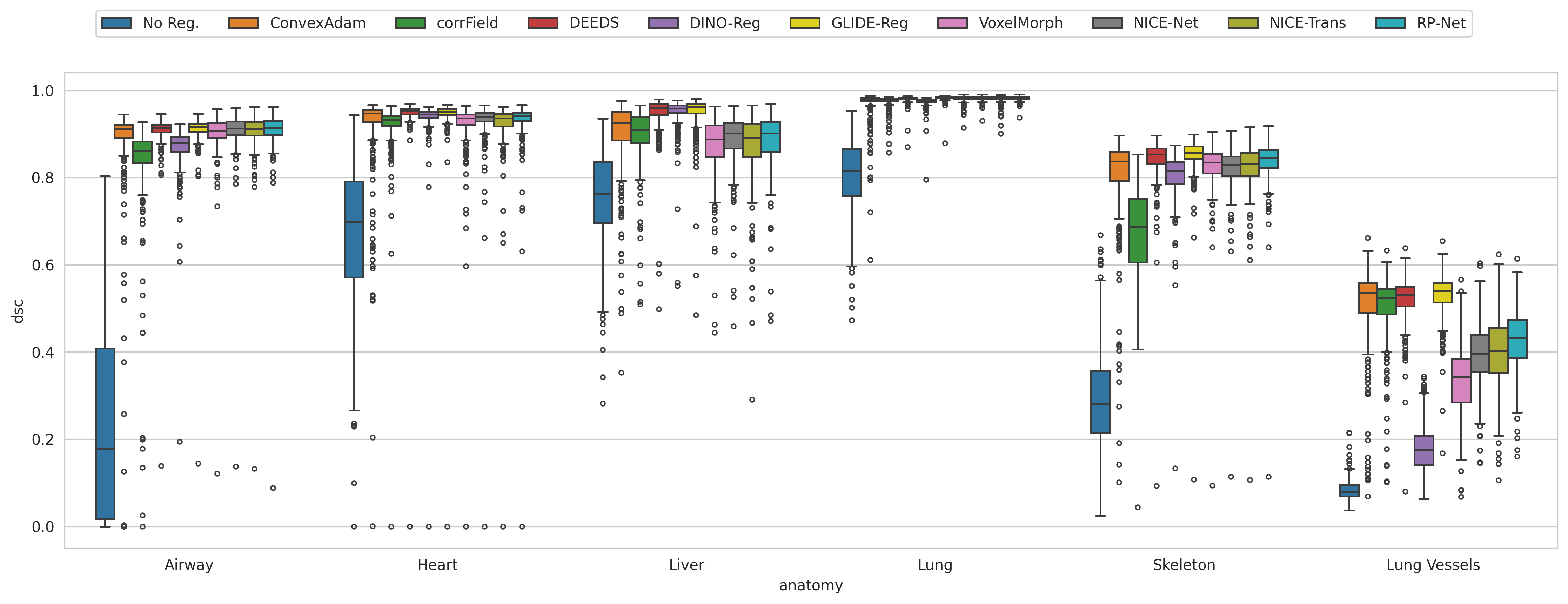}
\caption{Comparison of the best performing 9 algorithms (best four in the learning-based and all five feature-based instance-optimization algorithms) on the NLST dataset.} \label{fig:boxplot}
\end{figure}
\begin{figure}[htbp]
\includegraphics[width=\textwidth]{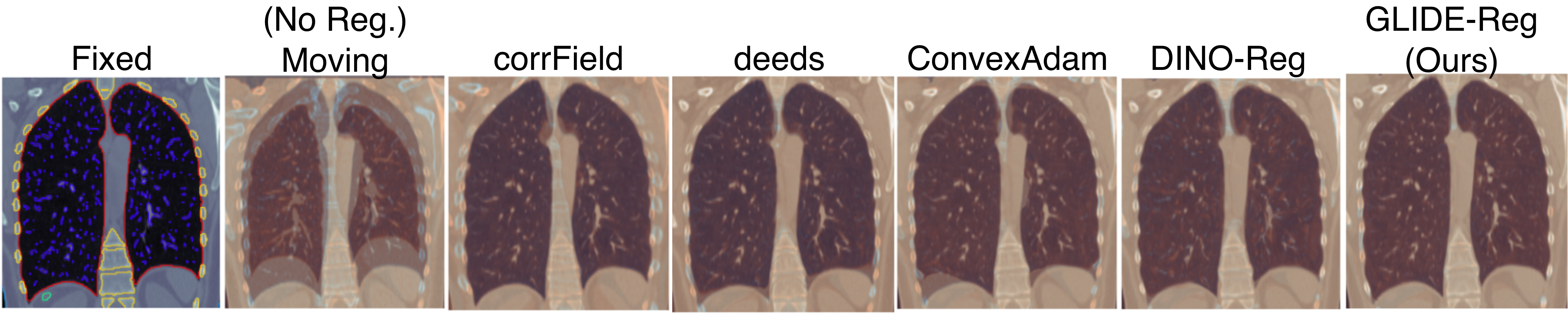}
\caption{Comparison of the feature-based instance-optimization methods on one challenging case from Lung250M, where all learning-based methods failed.} \label{fig:qualres}
\end{figure}
\subsection{Registration Results}
We compared the performance of our method GLIDE-Reg with learning-based benchmarks (VoxelMorph \cite{balakrishnan2019voxelmorph}, TransMorph \cite{chen2022transmorph}, ULAE-Net \cite{shu2021medical}, NICE-Net \cite{meng2022non}, NICE-Trans \cite{meng2023non}, and RP-Net \cite{ma2024iirp}) as well as feature-based benchmarks (corrField \cite{heinrich2015estimating}, DEEDS \cite{6471238}, ConvexAdam \cite{siebert2024convexadam}, and DINO-Reg \cite{song2024dino,song2025dino}). GLIDE-Reg outperforms all benchmark methods in terms of average DSC across all six anatomical structures. When comparing DSC across anatomical structures (Fig. \ref{fig:boxplot}), GLIDE-Reg improves most significantly on finer structures (e.g., airways, lung vessels). For $\mathbf{\%\lvert J \rvert_{<0}}$, GLIDE-Reg outperformed most learning-based algorithms and was comparable to the instance-optimized algorithms.

We also conducted a runtime comparison across two datasets (Lung250M and NLST) due to a significant difference in image resolution (Sec. \ref{sec:data}). 
Although corrField is faster, it performs poorly on anatomical structures that lack point features (e.g., any structure exterior to the provided lung mask). DEEDS has the second-best DSC and optimal nodule alignment, but suffers from a runtime that is twice that of GLIDE-Reg.

A challenging Lung250M case in Fig. \ref{fig:qualres} shows that GLIDE-Reg has the best matching in the overlaid image, whereas corrField suffers from skeleton alignment issues, and DINO-Reg misses the alignment of the smaller structures, such as the lung vessels. It is worth noting that learning-based methods perform poorly on this dataset because models' generalizability is limited when adapting to a relatively challenging dataset.

\begin{table}[htbp]
\begin{minipage}[t]{0.48\textwidth}
  \centering
  \caption{Ablation Study on dimensionality reduction methods and global-to-local module on Lung250M.}
  \resizebox{0.85\textwidth}{!}{
  \begin{tabular}{ccccc}
  \toprule
  \bfseries Method & 
  \bfseries TRE $\downarrow$ & \bfseries DSC $\uparrow$ & \bfseries Runtime $\downarrow$ \\
   & Landmarks & Mean6
  \\
  \midrule
  PCA & 
  $1.75 \pm 0.60$\textsuperscript{*} &
  $0.858 \pm 0.014$ & 
  $< 1.5$ min \\
  
  SDR & 
  $1.68 \pm 0.62$\textsuperscript{*} &
  $0.845 \pm 0.017$\textsuperscript{*} & 
  $< 1.5$ min \\
  
  %
  \midrule
  Global-only &
  $2.13 \pm 0.63$\textsuperscript{*} & 
  $0.840 \pm 0.017$\textsuperscript{*} &
  $< 1.5$ min \\
  Local-only & 
  $1.88 \pm 0.60$\textsuperscript{*} &
  $0.843 \pm 0.018$\textsuperscript{*} &
  $<10$ sec \\
  \midrule
  DDR/G2L &
  $\mathbf{1.58 \pm 0.50}$ & 
  $\mathbf{0.859 \pm 0.013}$ &
  $< 1.5$ min \\
  \bottomrule
  \end{tabular}
  }
  \label{tab:ablationddr}
\end{minipage}
\hfill
\begin{minipage}[t]{0.48\textwidth}
\centering
\caption{Ablation Study on nodule tracking with different CPM thresholds on NLST.}
\resizebox{0.85\textwidth}{!}{
  \begin{tabular}{cccccc}
  \toprule
  
  \bfseries Method & 
  \bfseries $0.5 mm$ & \bfseries $1.0 mm$ & \bfseries $2.0 mm$ & \bfseries $5.0 mm$
  \\
  \midrule
  corrField & 
  $40$ & $79$ & $90$ & $94$ \\
  DEEDS &
  $40$ & $75$ & $90$ & $\mathbf{97}$ \\
  ConvexAdam &
  $44$ & $73$ & $83$ & $89$ \\
  RP-Net & 
  $18$ & $42$ & $67$ & $84$ \\
  NICE-Trans & 
  $10$ & $32$ & $57$ & $85$ \\ 
  GLIDE-Reg &
  $\mathbf{46}$ & $\mathbf{81}$ & $\mathbf{92}$ & $\mathbf{97}$ \\
  \bottomrule  \\
  \end{tabular}
  }
  \label{tab:ablationnt}
\end{minipage}
\end{table}

\subsection{Ablation Study}
\textbf{Dynamic Dimensionality Reduction}. We first evaluated the effectiveness of different dimensionality reduction methods (Table \ref{tab:ablationddr}). Although PCA is also quite promising in DSC, TRE is the worst. VAE-based dimensionality reduction can achieve a TRE of $1.68 mm$ on static dimensionality reduction (SDR) with self-pretrained weights before registration optimization and $1.58 mm$ on dynamic dimensionality reduction (DDR) without pretraining.\\
\textbf{Global/Local-Only Registration}
We also evaluated the performance on global-only and local-only registration (Table \ref{tab:ablationddr}) and showed that GLIDE-Reg (G2L) registration outperforms the global/local-only registrations significantly.\\
\textbf{Interpretation of Nodule Tracking}
Table \ref{tab:ablationnt} reports the center point matching (CPM), which measures the percentage of predicted and reference lesion center points within $(0.5/1.0/2.0/5.0) mm$. GLIDE-Reg is consistently better than five other SOTA methods. 

\section{Conclusion}
We presented GLIDE-Reg, a registration algorithm that combines global and local feature extraction with coupled local and global registration within an instance-specific optimization framework. Additionally, we proposed VAE-based dimensionality reduction to better preserve semantically rich VFM features and jointly optimize registration displacement, thereby improving practicality and adaptability across multiple registration cases or tasks. 


%
%
%
\bibliographystyle{splncs04}
\bibliography{mybibliography_arxiv}
%




\end{document}